\documentclass[]{aa} 

\newcommand\lsim{\lower0.5ex\hbox{$\; \buildrel < \over \sim \;$}}
\newcommand\gsim{\lower0.5ex\hbox{$\; \buildrel > \over \sim \;$}}

\usepackage{graphicx}
\authorrunning{Dai \& Li}
\titlerunning{Accretion Torque on Magnetized Neutron Stars}

\begin{document}
   \title{Accretion torque on magnetized neutron stars}

   \author{Hai-Lang Dai and Xiang-Dong Li
          }

   \offprints{Xiang-Dong Li}

   \institute{Department of Astronomy, Nanjing University,
               Nanjing 210093, P. R. China\\
              \email{hldai@nju.edu.cn, lixd@nju.edu.cn}
                        }


   \abstract{
   The conventional picture of disk accretion onto magnetized neutron
stars has been challenged by the spin changes observed in a few
X-ray pulsars, and by theoretical results from numerical
simulations of disk-magnetized star interactions. These indicate
possible accretion during the propeller regime and the spin-down
torque increasing with the accretion rate. Here we present a model
for the accretion torque exerted by the disk on a magnetized
neutron star, assuming accretion continues even for rapid
rotators. The accretion torque is shown to have some different
characteristics from that in the conventional model, but in accord
with observations and numerical calculations of accretion-powered
magnetized neutron stars. We  also discuss its possible
applications to the spin evolution in X-ray pulsars.

   \keywords{ accretion, accretion disk --- pulsars: general
---stars: magnetic field---X-ray: stars
               }
   }

   \maketitle
%

\section{Introduction}

The interaction between a magnetized, rotating star and a
surrounding accretion disk takes place in a variety of
astrophysical systems, including T Tauri stars, cataclysmic
variables, and neutron star X-ray binaries (Frank, King, \& Raine
\cite{fra02}). Magnetic fields play an important role in
transferring angular momentum between the accreting star and the
disk. A detailed model for the disk-star interaction was developed
by Ghosh \& Lamb (\cite{gho79a}, \cite{gho79b}). In this model the
stellar magnetic field lines are assumed to penetrate the
accretion disk, and become twisted because of the differential
rotation between the star and the disk. The resulting torque can
spin up or down the central star, depending on the star's
rotation, magnetic field strength, and mass accretion rate. The
total torque $N$ exerted on a star with mass $M$ contains two
components,
\begin{equation}
N=N_0+N_{\rm mag}.
\end{equation}
The torque $N_0$ carried by the material falling onto the star is
generally taken to be
\begin{equation}
N_0=\dot{M}R_0^2\Omega_0
\end{equation}
where $\Omega_0\simeq (GM/R_0^3)^{1/2}$ is the (Keplerian) angular
velocity at the inner radius $R_0$ of the disk, $G$ the
gravitational constant, and $\dot{M}$ the accretion rate. The
magnetic torque $N_{\rm mag}$ results from the azimuthal field
component $B_{\phi}$ generated by shear motion between the disk
and the vertical field component $B_{\rm z}$\footnote{Here we have
adopted cylindrical coordinate $(R, \phi, z)$ centered on the
star, and the disk is assumed to be located on the $z=0$ plane,
perpendicular to the star's spin and magnetic axes.},
\begin{equation}
N_{\rm mag}=-\int_{R_0}^{\infty}R^2B_{\phi}B_{\rm z}dR.
\end{equation}
The torque $N_{\rm mag}$ can be positive and negative, depending
on the value of the ``fastness parameter" $\omega=\Omega_{\rm
*}/\Omega_0=(R_0/R_{\rm c})^{3/2}$, where $\Omega_{\rm *}$ is the
angular velocity of the star and $R_{\rm c}\equiv(GM/\Omega_{\rm
*}^2)^{1/3}$ the corotation radius. A rapidly rotating star
($\omega\lsim 1$) can be spun down with steady accretion. The
Ghosh \& Lamb picture has been modified and extended in several
directions by, for example, K$\ddot{o}$nigl (\cite{k91}), Campbell
(\cite{cam92}), Lovelace et al. (\cite{lov95}), Wang
(\cite{wang95}), Yi (\cite{yi95}), Li \& Wang (\cite{lw96}), Li,
Wickramasinghe, \& R$\ddot{u}$diger (\cite{lwr96}).

However, the above standard model has been challenged by recent
observations on the spin changes in several X-ray pulsars
(Bildsten et al. \cite{bil97}) and accreting millisecond pulsars
(Galloway et al. \cite{gal02}), and by theoretical results from 2-
and 3-dimensional numerical simulations of disk accretion to
magnetized, rotating stars (Romanova et al. \cite{rom02},
\cite{rom03}, \cite{rom04}). These observational and theoretical
facts have raised a number of puzzling issues, including  (1)
abrupt spin reversals in X-ray pulsars, (2) spin-down in
millisecond X-ray pulsars throughout the outburst, (3) spin-down
rates increasing with accretion rate, and (4) accretion in the
``propeller" regime. These have motivated more recent
investigations on the torques exerted by an accretion disk on a
magnetized star (Nelson et al. \cite{nel97}; Yi, Wheeler, \&
Vishniac \cite{yi97}; Torkelsson \cite{tor98}; Lovelace, Romanova,
\& Bisnovatyi-Kogan \cite{lov99}; Locsei \& Melatos \cite{loc04};
Rappaport et al. \cite{rap04}).

In this paper, attempting to solve some of the puzzles related to
disk-accreting neutron stars, we have constructed a new model for
the accretion torque based on some modifications of the Ghosh \&
Lamb picture. We introduce the model in section 2, and apply it to
spin evolution in X-ray pulsars in section 3. In section 4 we
discuss the similarities and differences between this work and the
competent model proposed by Lovelace et al. (\cite{lov99}).

\section{Model}

Here we adopt the typical assumptions. The central star contains a
rotation-axis aligned dipolar magnetic field. At a cylindrical
radius $R$ from the star, the vertical component of the field
$B_{\rm z}(R)=B_*(R_*/R)^3$, where $B_*$ is the stellar surface
field strength. This field is anchored in the stellar surface and
also threads a thin-Keplerian accretion disk. The region above and
below the disk contains low-density gas and the system exists in a
steady-state. Assume that the magnetosphere is nearly force-free
and reconnection takes place outside the disk (Wang
\cite{wang95}), the azimuthal component $B_{\phi}$ of the field on
the surface of the disk generated by rotation shear is given by
\begin{equation}
\frac{B_{\phi}}{B_z}=\left\{
\begin{array}{l l}
\gamma\frac{(\Omega_*-\Omega_{\rm K})}{\Omega_{\rm K}}, & \Omega_*\leq\Omega_{\rm K} \\
\gamma\frac{(\Omega_*-\Omega_{\rm K})}{\Omega_*}, &
\Omega_*>\Omega_{\rm K},
\end{array}
\right.
\end{equation}
where the parameter $\gamma$ is of order unity, and $\Omega_{\rm
K}=\Omega_{\rm K}(R)$ is the Keplerian angular velocity in the
disk. This form guarantees that the azimuthal pitch
$|B_{\phi}/B_{\rm z}|$ cannot be larger than $\gamma$ over
significant distances, for reasons of equilibrium and stability of
the field above and below the disk plane (Aly \cite{aly84},
\cite{aly85}; Uzdensky, K$\ddot{o}$nigl, \& Litwin \cite{uz02}).

We assume that Eq.~(4) applies in both accretion ($\omega < 1$)
and ``soft" propeller ($\omega\gsim 1$) cases. It is
conventionally thought that, when $\dot{M}$ drops sufficiently, an
accreting neutron star will transit from an accretor to a
propeller, in which mass ejection occurs. However, this point of
view is not fully consistent with recent observations of
transient, accretion-driven X-ray pulsars and with the energetics
considerations (see Rappaport et al. \cite{rap04} for a
discussion). In stead, Spruit \& Taam (\cite{spr93}) argued that
even when $\omega \gsim 1$ accretion cannot stop, and the disk may
find a new accretion state with the inner disk radius close to the
corotation radius $R_{\rm c}$. Rappaport et al. (\cite{rap04})
further developed this idea and applied the results to compute the
torques expected during the outbursts of the transient millisecond
pulsars.

The inner edge $R_0$ of an magnetically truncated accretion disk
is usually assumed to be determined by the condition that the
magnetic stresses dominate over viscous stresses in the disk and
begin to disrupt the Keplerian motion of the disk material
(Campbell \cite{cam92}; Wang \cite{wang95}), i.e.,
\begin{equation}
-R_0^2B_{\phi0}B_{z0}=\dot{M}\frac{{\rm d}(R^2\Omega)}{{\rm d}
R}|_0,
\end{equation}
where the subscript 0 denotes quantities evaluated at $R_0$. The
limitation of this definition is obvious if one combines Eqs. (4)
and (5), and takes the limit as $\Omega$ approaches $\Omega_0$:
the left hand side of Eq.~(5) becomes zero while the right hand
side keeps to be a certain value, suggesting that it is not
possible to calculate $R_0$ when $\omega\rightarrow 1$ (Li \&
Wickramasinghe \cite{lw97}). However, the critical fastness
parameter $\omega_{\rm c}$ beyond which spin-down occurs is always
close to 1 (Wang \cite{wang95}, \cite{wang97}; Li \& Wang
\cite{lw96}). Since in our model accretion occurs with $\omega$
extending from 0 to $\gsim 1$, we in stead take $R_0$ to be at the
magnetospheric radius $R_{\rm m}$ determined by equating the ram
pressure of the accreting matter with the magnetic pressure due to
the dipole field of the neutron star (Davidson \& Ostriker
\cite{dav73}; Lamb, Pethick, \& Pines \cite{lam73}),
\begin{equation}
R_0\simeq R_{\rm m}=(\frac{\mu^4}{2GM\dot{M}^2})^{1/7},
\end{equation}
where $\mu=B_*R_*^3$ is the magnetic moment of the neutron star.
Observational evidence supporting this expression has been found
in the transient X-ray pulsar A0535+26 (Li \cite{li97}). We have
not taken $R_0=R_{\rm c}$ for $\omega>1$ as in Rappaport et al.
(\cite{rap04}). The reason is that, if the disk has already been
disrupted at $R_{\rm m}$ $(>R_{\rm c})$, the accretion flow may
pile up around the magnetosphere (Romanova et al. \cite{rom04}),
it is unknown whether the accretion flow can still keep the
``disk" structure between $R_{\rm c}$ and $R_{\rm m}$.

We now estimate the torque exerted on the neutron star by the
surrounding accretion disk. When the disk rotation is Keplerian,
we integrate the magnetic torque in Eq.~(3) over the disk from
$R_0$ to infinity combining with Eq.~(4), and get the following
expression,
\begin{eqnarray}
N_{\rm mag}&=&\left\{
 \begin{array} {l l}
(\frac{\gamma\mu^2}{3R_0^3})(1-2\omega+\frac{2\omega^2}{3}), & \omega\leq 1
\\
(\frac{\gamma\mu^2}{3R_0^3})(\frac{2}{3\omega}-1), & \omega>1.
\end{array}
\right.
\end{eqnarray}

For the material torque $N_0$, Eq.~(2) implies 100\% efficient
angular momentum transfer. It actually results from the
magnetospheric torque arising from the shearing motion between the
corotating magnetosphere and the non-Keplerian boundary layer in
the disk, which connects the magnetosphere and the outer Keplerian
disk. The multi-dimensional calculations by Romanova et al.
(\cite{rom02}, \cite{rom04}) indicate that most of the angular
momentum carried by the accreting material in the magnetosphere
has been transferred to the magnetic field. From the
investigations of magnetized accretion disks (e.g., Ghosh \& Lamb
\cite{gho79a}; Campbell \cite{cam87}; Erkut \& Alpar \cite{ea04})
we find that the angular velocity in the boundary layer can be
roughly represented with the following form
\begin{equation}
\Omega(R)=\Omega_{0}(R/R_0)^n,
\end{equation}
and the magnetospheric torque can be evaluated according to
Eq.~(4),
\begin{eqnarray}
N_{0}& = & -\int_{R_0-\Delta}^{R_0}R^2B_{\phi}B_{\rm
z}dR \nonumber \\
 & = & \gamma\mu^2\int_{R_0-\Delta}^{R_0}R^{-4}[1-(\frac{\Omega_*}{\Omega_0})
      (\frac{R}{R_0})^{-n}]dR \nonumber \\
 & \simeq & \gamma\delta\frac{\mu^2}{R_0^3}(1-\omega) \nonumber \\
 & = & (\sqrt{2}\gamma\delta)\dot{M}(GMR_0)^{1/2}(1-\omega)
\end{eqnarray}
where $\Delta$ is the width of the boundary layer and
$\delta=\Delta/R_0\ll 1$ (Ghosh \& Lamb \cite{gho79a}).

Obviously the real form of $\Omega(R)$ must be more complicated
than Eq.~(8). However, Eq.~(9) presents a reasonable
order-of-magnitude estimate of the magnetospheric torque. Thus we
let the torque $N_0$ take the following form in both accretion and
propeller cases (see also Menou et al. \cite{men99}),
\begin{equation}
N_0=\xi\dot{M}(GMR_0)^{1/2}(1-\omega),
\end{equation}
where the parameter $\xi=\sqrt{2}\gamma\delta$ depends on the
structure of the magnetosphere and $0< \xi\lsim 1$. This implies
that {\em the accreting material in the boundary layer may lose
some fraction of the angular momentum, probably, through the
viscous torque and/or the magnetic centrifugal winds/outflows}
(e.g. Erkut \& Alpar \cite{ea04}). In the present analysis we
always take $\xi=1$, and Eq.~(10) recovers to Eq.~(2) when
$\omega\ll1$. The total torque exerted on the star by the disk is
then
\begin{eqnarray}
N&=&N_0+N_{\rm mag} \nonumber\\
 &=&\left\{
 \begin{array} {l}
\dot{M}(GMR_0)^{1/2}[\xi(1-\omega)
+\frac{\sqrt{2}\gamma}{3}(1-2\omega+\frac{2\omega^2}{3})],
\ \omega\leq 1 \\
\dot{M}(GMR_0)^{1/2}[\xi(1-\omega)
 +\frac{\sqrt{2}\gamma}{3}(\frac{2}{3\omega}-1)],
\  \omega>1.
\end{array}
\right.
\end{eqnarray}

Figures 1a and 1b show the dependence of the dimensionless torque
$N/\dot{M}(GMR_0)^{1/2}$ and $N/(\mu^2/\sqrt{2}R_c^{3})$ on
$\omega$ in solid curves with $\xi=\gamma=1$, respectively. The
dashed curves in the figures represent the results in the disk
model of Wang (\cite{wang95}) for comparison. In Fig.~1a there is
no singularity problem in this work when $\omega\rightarrow 1$,
the parameter space range for $\omega$ for spin-down with
accretion is also much wider. The critical fastness parameter
$\omega_{\rm c}$, at which $N=0$, is $\omega_c\simeq 0.884$.
Furthermore, it indicates that, for constant $\dot{M}$, the
spin-down torque increases with $\Omega_*$ ($\propto\omega$), in
line with the numerical simulation results by Romanova et al.
(\cite{rom04}).

Figure 1b shows how the torque varies with
$\omega\propto\dot{M}^{-3/7}$ when $\mu$ and $\Omega_*$ are
invariant. One can see that the spin-down torque is not a
monotonous function of $\omega$ (or $\dot{M}$) when
$\omega>\omega_{\rm c}$.  The spin-down torque takes the maximum
value when $\omega\simeq 1.634$. Thus a given spin-down torque can
correspond to two values of $\omega$. {\em When $\omega>1.634$,
the spin-down torque increases with $\dot{M}$}, which is opposite
to the dependence in the standard model, but consistent both with
the observational fact in the X-ray pulsar GX 1+4 that the X-ray
flux appears to be increasing with the spin-down torque
(Chakrabarty \cite{cha95}), and with the numerical simulation
results by Romanova et al. (\cite{rom04}).

Finally we point out that the derived results in this simplified
model should be regarded as qualitative, and one should not take
the values above very seriously.

\section{Applications to X-ray pulsars}

The long-term, continuous monitoring of accreting X-ray pulsars by
{\em BATSE} on board {\em Compton Gamma-Ray Observatory} has
provided new insight into the spin evolution of these systems
(Bildsten et al. \cite{bil97}). For the X-ray pulsar Cen X$-$3,
the {\em BATSE} data show that it exhibits 10$-$100 day intervals
of steady spin-up and spin-down, with a transition timescale
$\lsim 10$ days. Similar bimodal torque (or torque reversal)
behavior has been observed in other pulsars, including 4U
1626$-$67, GX 1+4, and OAO 1657$-$415. These pulsars seem to be
subject to instantaneous torques of roughly comparable magnitude
$\lsim \dot{M}(GMR_{\rm c})^{1/2}$, and differentiate themselves
only by the timescale for reversals of sign. It is hard for the
standard Ghosh \& Lamb type model to explain a sudden torque
reversal with nearly constant $|N|$ unless the $\dot{M}$ variation
is discontinuous and fine tuned.

Many mechanisms have been proposed to explain the torque reversals
in accretion-powered X-ray pulsars. They can be roughly divided
into two categories, one for the change of the direction of disk
rotation, the other for bimodal magnetic torque. In the former
type of models, van Kerkwijk et al. (\cite{van98}) suggested that
the accretion disk may be subject to a warping instability because
of the irradiation from the neutron star. The inner part of the
accretion disk will flip over and rotate in the opposite
direction, which would lead to a torque reversal. The spin-down
can also be explained if the disk rotation is retrograde
(Makishima et al. \cite{mak88}; Nelson et al. \cite{nel97}). In
the latter, Torkelsson (\cite{tor98}) suggested that the magnetic
torque reversals may be explained by the presence of the intrinsic
magnetic field in the accretion disk. The orientation of the disk
field is not determined by the difference between the angular
velocities of the star and of the disk but is rather a free
parameter. Thus the direction of the magnetic torque between the
two is arbitrary. A bimodal magnetic torque was also proposed by
Lovelace et al. (\cite{lov99}, see below). Locsei \& Melatos
(\cite{loc04}) presented a model of the disk-magnetosphere
interaction, which adds diffusion of the stellar magnetic field to
the disk. In certain conditions, the system possesses two stable
equilibria, corresponding to spin-up and spin-down.

An alternative model for the transitions between spin-up and
spin-down was suggested by Yi et al. (\cite{yi97}), in which the
torque reversals are caused by alternation between a Keplerian
thin disk and a sub-Keplerian, advection-dominated accretion flow
(ADAF) with small changes in the accretion rate. When $\dot{M}$
becomes smaller than a critical value $\dot{M}_{\rm cr}$, the
inner part of the accretion disk may make a transition from a
primarily Keplerian flow to a substantially sub-Keplerian ADAF, in
which the angular velocity $\Omega'(R)=A\Omega_{\rm K}(R)$ with
$A<1$ (Narayan \& Yi \cite{nara95}). In this case the corotation
radius becomes $R'_{\rm c}=A^{2/3}R_{\rm c}$ (Here we use the
prime to denote quantities in ADAF). The dynamical changes in the
disk structure lead to the slow ($\omega<\omega_c$, spin-up) and
rapid ($\omega'>\omega'_c$, spin-down) rotator stage alternatively
with $\dot{M}\sim\dot{M}_{\rm cr}$. This scenario is attractive
because it is consistent with the transition time scales - the
standard disk-ADAF transition will occur on a thermal time
$t_{th}\sim (\alpha\Omega_0)^{-1}\sim 10^3$ s, while the interval
between torque transition is set by the changes in $\dot{M}$ on a
global viscous time scale. However, some issues in this model
suggest that it needs to be improved. First, at equilibrium spin
the critical fastness parameter $\omega_{\rm c}=\omega'_{\rm
c}\simeq 0.875$ (see also Wang \cite{wang95}). The narrow range of
$\omega'$ ($\sim 0.875-1$) for spin-down requires that the
physical parameters (e.g., $\dot{M}$, $B_*$) should still be
fine-tuned for various X-ray pulsars\footnote{For oblique
rotators, there could be no spin-down allowed with $\omega$ or
$\omega'<1$ (Wang \cite{wang97}).}. Second, as the Ghosh \& Lamb
model, this model also predicts decreased spin-down rate when
$\dot{M}$ increases.

In the present work, we have assumed that accretion can continue
even when $\omega\gsim 1$. During the disk transition the inner
radius of the disk may change little, i.e., $R'_0\simeq R_0$,
since the total (thermal and kinetic) energy density in the disk
remains to be unchanged (Yi et al. \cite{yi97}). The torque in the
ADAF case can be derived in the similar way as in the Keplerian
disk case,
\begin{eqnarray}
N'&=&\left\{
 \begin{array} {l}
A\dot{M}(GMR'_0)^{1/2}[\xi(1-\omega')+\frac{\sqrt{2}\gamma}{3A}
(1-2\omega'+\frac{2\omega'^2}{3})],
\ \omega'\leq 1 \\
A\dot{M}(GMR'_0)^{1/2}[\xi(1-\omega')+\frac{\sqrt{2}\gamma}{3A}
(\frac{2}{3\omega'}-1)], \ \omega'>1
\end{array}
\right.
\end{eqnarray}
where $\omega'=(R_0'/R_{\rm c}')^{3/2}=\omega/A$. Note that the
critical fastness parameter $\omega'_{\rm c}$ (for $N'=0$) equals
$\omega_c$, but the equilibrium period would become longer by a
factor of $1/A$, and the system begins to evolve towards the newly
determined equilibrium after the transition. Figure 2 shows the
torque $N$ and $N'$ in units of $\dot{M}(GMR_{\rm c})^{1/2}$ as a
function of $\omega$ with $\xi=\gamma=1$. For $\omega$ between
$\sim 0.2$ and $\sim 0.8$, if $\dot{M}$ is around $\dot{M}_{\rm
cr}$, small changes in $\dot{M}$ can cause the torque reversal
behavior with comparable spin-up/down torques
$\lsim\dot{M}(GMR_{\rm c})^{1/2}$.

We have applied the model to explain the spin change in several
X-ray pulsars. As an illustration, Fig.~3 shows the fit to the
observed spin reversal in GX 1+4. Similar as in Yi et al.
(\cite{yi97}), we assume that $\dot{M}$ keeps a linear increase or
decrease as an approximation to more complex $\dot{M}$ variations
on longer timescales. The transition occurs on a timescale $ \ll
|\Omega_*/\dot{\Omega}_*|$ before and after the transition. In
Fig.~3, the observed torque reversal event is reproduced by taking
$B_*=1.0\times10^{13}$ G, $\dot{M}=6.0\times10^{15}$ gs$^{-1}$,
$d\dot{M}/dt=3.25\times10^{15}$ gs$^{-1}$ yr$^{-1}$, and $A=0.3$
(we take the neutron star mass $M=1.4 M_{\odot}$, and radius
$R_*=10^6$ cm). We note, however, that our estimated values (of
$B_*$ and $\dot{M}$) can always be rescaled by changing $A$  and
$\gamma$. The critical accretion rate for the transition is
assumed to be $\dot{M}_{\rm cr}=6.5\times10^{15}$ gs$^{-1}$.

\section{Discussion}
Assuming accretion continues during the (soft) propeller stage, we
have constructed a model for the magnetized torque on accreting
neutron stars, which changes continuously when $\omega$ varies
from 0 to $\gsim 1$ (without the singularity problem), and the
spin-down torque increases with $\dot{M}$. Following Yi et al.
(\cite{yi97}), we have applied the model to explain the torque
reversal phenomena observed in a few X-ray pulsars.

The same problem has been investigated by Lovelace et al.
(\cite{lov99}). These authors have developed a model for magnetic
``propeller"-driven outflows for rapidly rotating neutron stars
accreting from a disk. They find that the inner radius $R_0$ of
the disk depends not only on $\mu$ and $\dot{M}$, but also on the
star's rotation rate $\Omega_*$. The most interesting result is
that for a given value of $\Omega_*$, there may exist two
solutions of $R_0$, with one $>R_{\rm c}$ and the other $<R_{\rm
c}$, corresponding to two possible equilibrium configurations. In
a transition of $R_0$, the propeller varies between being ``off"
and being ``on", leading to jumps between spin-up and spin-down.
The transitions are assumed to be stochastic or chaotic in nature,
and could be triggered by small variations in the accretion flow
and magnetic field configuration. The ratio of spin-down to
spin-up torque is also shown to be of order unity.

This model shows some similar features as our work. For example,
in both models (1) discontinuous transitions of the magnetic
torque are invoked to account for the observed abrupt spin changes
in X-ray pulsars, (2) the torques are of comparable magnitude, and
(3) the spin-down torques increases with $\dot{M}$. Especially,
Eq.~(10) suggests loss of angular momentum at the inner edge of
the disk, probably by the magnetic-driven outflows proposed by
Lovelace et al. (\cite{lov99}).

However, there exist significant differences between the two
works. First, in the propeller regime, Lovelace et al.
(\cite{lov99}) let most of the accreting material be ejected from
the star, while we still allow mass accretion to guarantees no
significant variations in the X-ray luminosities during the
transition of spin evolution, implying that $\omega\gg1$ in
Lovelace et al. (\cite{lov99}), and $\omega\gsim 1$ in this work.
Second, the torque reversals in our model are caused by the change
of the dynamical structure of the disk, as suggested by Yi et al.
(1997), at a certain, critical mass accretion rate. In Lovelace et
al. (\cite{lov99}) the transition is stochastic, and could take
place at any luminosities. Third, we have adopted the traditional
Alf\'ven radius as the inner radius $R_0$ of the disk in both
accretion and propeller phases. In Lovelace et al. (1999) $R_0$
depends on $\Omega_*$ also, and its value increases with
$\Omega_*$ in the propeller case. Obviously the hypothesis of
$\Omega_*$-depending $R_0$ is more realistic, but how $R_0$
changes with $\Omega_*$ has not been well understood\footnote{
Romanova et al. (\cite{rom04}) indeed observed increasing $R_0$
with $\Omega_*$ in their calculations. But this change is caused
by the fact that the initial dipolar field becomes non-dipolar
when the star rotates very rapidly.}. Future observations are
expected to provide more stringent constraints on the mechanisms
suggested in both models.

\begin{acknowledgements}
This work was supported by NSFC under grant number 10573010.
\end{acknowledgements}

\clearpage
\begin{figure*}
\centering
  \includegraphics[width=17cm]{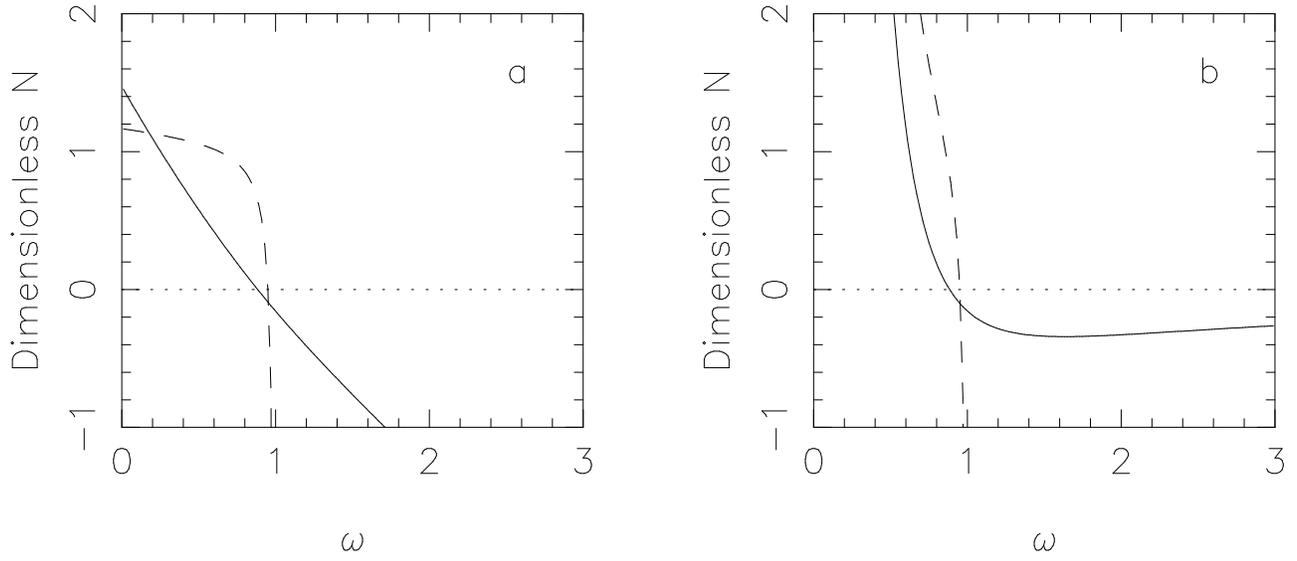}
  \caption{The solid lines show the dimensionless torque (a)
  $N/\dot{M}(GMR_0)^{1/2}$ and (b) $N/(\mu^2/3R_{\rm c}^3)$ as a
  function of the fastness parameter $\omega$. The dashed lines
  represent the standard model. }
\end{figure*}

\clearpage
\begin{figure}
\resizebox{\hsize}{!}{\includegraphics{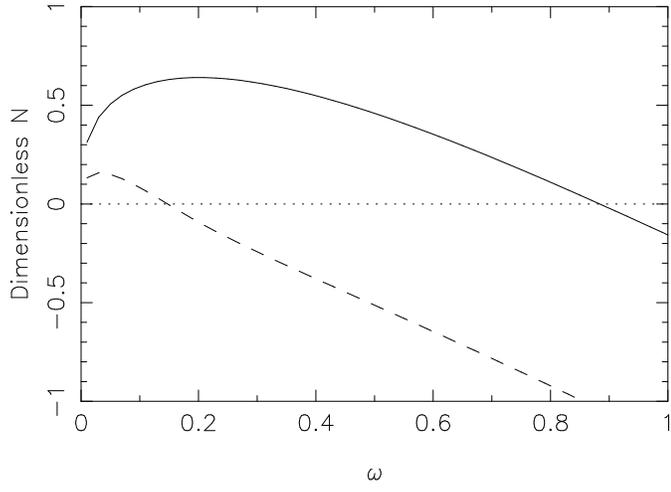}} \caption{The
dimensionless torque $N/\dot{M}(GMR_{\rm c})^{1/2}$ (solid line)
and $N'/\dot{M}(GMR_{\rm c})^{1/2}$ (dashed line) as a function of
the fastness parameter $\omega$.}
\end{figure}

\clearpage
\begin{figure}
\resizebox{\hsize}{!}{\includegraphics{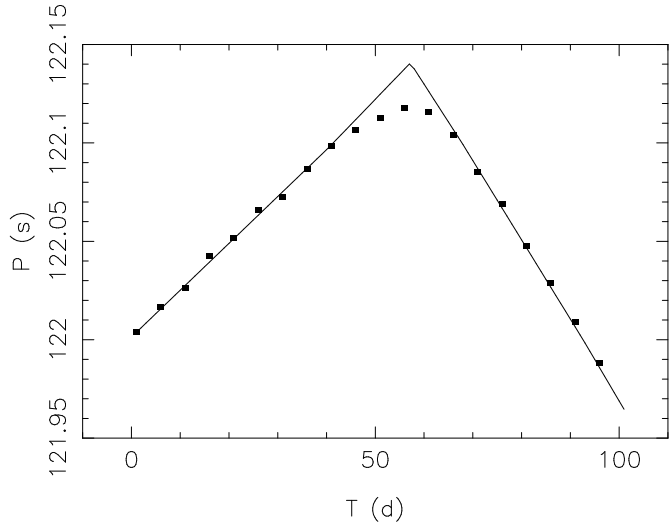}}
\caption{Torque reversal event in GX 1+4. The points are the
observed data and the solid line corresponds to the fitting
result. }
\end{figure}


\end{document}